\documentclass[final,5p,times,twocolumn,authoryear]{elsarticle}
\usepackage{amssymb}
\usepackage{lipsum}
\pdfoutput=1
\usepackage{tikz}
\usepackage{amsmath}
\usepackage{siunitx}
\usepackage{amsthm}
\usepackage{mathtools, cuted}
\usepackage{multicol}
\usepackage{multirow}

\journal{Physics Open}

\begin{document}

\begin{frontmatter}

\title{Bi-planar magnetic stabilisation coils for an inertial sensor based on atom interferometry}

\affiliation[UoN]{organization={School of Physics \& Astronomy},%Department and Organization
            addressline={University of Nottingham}, 
            city={Nottingham},
            postcode={NG7 2RD}, 
            country={United Kingdom}}

\affiliation[ICL]{organization={Department of Physics},%Department and Organization
            addressline={Imperial College London}, 
            city={London},
            postcode={SW7 2BW}, 
            country={United Kingdom}}

\author[UoN]{A. Davis}
\author[UoN]{P. J. Hobson}
\author[UoN]{T. X. Smith}
\author[UoN]{C. Morley}
\author[ICL]{H. G. Sewell}
\author[ICL]{J. Cotter}
\author[UoN]{T. M. Fromhold}
\ead{Mark.Fromhold@nottingham.ac.uk}

\begin{abstract} %% Text of abstract
Inertial sensors that measure the acceleration of ultracold atoms promise unrivalled accuracy compared to classical equivalents. However, atomic systems are sensitive to various perturbations, including magnetic fields, which can introduce measurement inaccuracies. To address this challenge, we have designed, manufactured, and validated a magnetic field stabilisation system for a quantum sensor based on atom interferometry. We solve for the magnetic field generated by surface currents in-between a pair of rectangular coils and approximate the surface current using discrete wires. The wires are wound by-hand onto machined panels which are retrofitted onto the existing mounting structure of the sensor without interfering with any experimental components. Along the central $60$~mm of the $y$-axis, which aligns with the trajectory of the atoms during interferometry, the coils are measured to generate an independent uniform axial magnetic field with a strength of $B_z=\left(22.81\pm0.01\right)$~$\mu$T/A [$\mathrm{mean}\pm2\sigma~\mathrm{std.~error}$] and an independent linear axial field gradient of strength $\mathrm{d}B_z/\mathrm{d}y=\left(10.6\pm0.1\right)$~$\mu$T/Am. The uniform $B_z$ field is measured to deviate by a maximum value of $1.3$\% in the same region, which is a factor of three times more uniform than the previously-used on-sensor rectangular $B_z$ compensation set.
\end{abstract}


\end{frontmatter}

\section{Introduction}\label{introduction}
Advances in the understanding and control of quantum phenomena have enabled devices that exploit quantum effects to become the state-of-the-art for many research and industrial applications. This extends from quantum computers to understanding protein folding~\citep{Robert_2021}, to quantum sensors [see review in~\cite{BongsNature}] like networked quantum clocks to improve accuracy of satellite navigation systems~\citep{K_m_r_2014}, ultra-precise atomic magnetometers for moving brain scanning~\citep{10.1038/nature26147}, and atom-interferometric sensors for gravity mapping~\citep{Wueaax0800,Stray2021} and inertial navigation~\citep{doi:10.1080/23746149.2021.1946426}. 

Most atom interferometers operate by applying sequence of light pulses which represent a quantum analogue of Mach--Zehnder interferometry to determine the acceleration or rotation of an atomic cloud~\citep{PhysRevA.65.033608}. By monitoring these forces over time, they can be used as inertial navigators which determine position relative to a reference point without requiring external input which may be intercepted or blocked. Typically, the interferometer phase is encoded in oscillations of the populations of atoms in two ground states and is proportional to the inertial forces experienced by the atoms~\citep{berman1997atom}. Significant research has focused on demonstrating atom interferometry in real-world settings~\citep{AIQuantumSensors}, including under the sea~\citep{10.1038/s41467-018-03040-2} and during flight~\citep{Bidel_2020}, and there are future projects to apply them for gravity measurements in space~\citep{Lachmann_2021,he2023space}. The resilience of sub-components of these systems is also a subject of extensive investigation. For example, Magneto--Optical Traps (MOTs), which are required to collect the atoms before interferometry, have been shown to operate during flight~\citep{atoms10010032} and even inside a near-surface borehole~\citep{10.1371/journal.pone.0288353}.

However, a limit to the widespread adoption of atom interferometers as inertial sensors is that errors are introduced by uncontrolled background magnetic field variations. These background variations may limit the atom temperature in MOTs~\citep{Lett:89} or manifest directly in the interferometer phase if magnetically sensitive ground state sub-levels are used for interferometry~\citep{PhysRevLett.117.023001}. If a strong bias field is used to shift magnetically-sensitive sub-levels off-resonance, then an additional systematic bias may be introduced due to higher-order corrections to the Zeeman shift~\citep{Wu_2014,HuZeeman} if the bias field is not sufficiently uniform. Background variations may be attenuated using layers of high permeability magnetic shielding~\citep{doi:10.1063/1.5141340,Hobson_2022}, but this increases the Size, Weight, Power, and Cost (SWaP-C) footprint of the sensor and needs to be physically-protected from impacts that may permanently magnetise the shielding. Instead, externally driven active coils can be used to simultaneously null variations and generate more uniform biasses. But, these active coils must be carefully designed and manufactured to balance their performance with ease of manufacture and durability in hostile environments.

Many methods to design these coils were pioneered by scientists developing more accurate components of magnetic resonance scanners. \cite{10.1088/0022-3727/19/8/001} introduced the target field method where a continuous current distribution on a cylindrical surface to generate a target magnetic field inside of it is found using Fourier inversion of a Green's function solution to Poisson's equation~\citep{JacksonEM}. \cite{yoda1990analytical} demonstrated that the target field method may be applied to design continuous current distributions on planar surfaces and \cite{10.1088/0022-3727/34/24/305} showed that surface current distributions may be solved as Fourier series, allowing simple optimisation techniques to be applied to determine the target field in terms of each mode on the surface. \cite{Pissanetzky_1992} introduced a numerical target field method to solve for continuum current distributions based on meshes that may be adapted to any geometry, at the cost of increased computational effort.

Recently, there has been extensive work applying target field methods to design the ultra-precise magnetic fields required for quantum technologies. For example, \cite{Holmes2019} used symmetry arguments to generate an accessible design formulation for rectangular bi-planes to control magnetic fields around atomic magnetometers and \cite{PhysRevApplied.18.014036} applied a numerical implementation of a target field method to design precise on-board circuits for an atomic magnetometer. Various target field solutions have also been posed which couple directly to the distortion introduced by high-permeability magnetic shielding~\citep{PhysRevApplied.15.054004,PhysRevApplied.15.064006,10.1063/5.0151057}, allowing the development of high-performance coupled shielding environments for atomic magnetometers~\citep{10177829}. A critical consideration during the design process is the effectiveness to which the continuous current can be approximated using discrete wires~\citep{10.1016/j.jmr.2005.07.003,DiscretePaper}: more wires allows the continuum solution to be approximated better but makes the coil harder to fabricate. There are also multiple fabrication options, some of which allow extensive geometric flexibility including 3D-printing coil formers~\citep{Hobson_2022} or rigid and flexible Printed Circuit Boards (PCBs)~\citep{10177829,PhysRevApplied.18.014036}, but this must be balanced against high costs as circuit size increases.

Here, we adapt the analytic coil design formulation posed by \cite{Yoda1990} to design a magnetic stabilisation system which is retrofitted to an inertial sensor based on atom interferometry. We pose the design model to exploit symmetries within the current flow on the planar surfaces, so that we only optimise the specific contributions that generate the target fields. We design two independent coil systems to stabilise two magnetic field variations which are critical for sensor operation: a uniform axial field and a linear gradient in the axial field with respect to the trajectory of the atoms during interferometry. The coil systems are manufactured using Computer Numerical Control (CNC) routing of wire paths onto pairs of acrylic sheets of area $570$ x $762$ mm$^{2}$, which are then hand-wound with enameled copper wire. The sheets in each pair are mounted on opposite sides of the interferometer, directly integrated into its mounting frame to minimise their spatial footprint. Their performance is validated experimentally and compared to the analytic design model. Finally, the future prospects for on-board implementation of the system onto the sensor and minimisation of its SWaP-C footprint are explored.

\section{Theory} \label{sec:theory}

\subsection{The inertial sensor}
Quantum-enhanced inertial sensors use coherent superpositions of matter waves to measure the acceleration or rotation-rate of a platform with an extremely low measurement bias and bias drift. Typically, atoms are cooled in a MOT and then an optical molasses to reach a temperature of a few $\mu$K. Atoms are then prepared in a single hyperfine state which is insensitive to first-order Zeeman shift. Two-photon stimulated Raman transitions then split, reflect, and recombine the atomic wave packets, in a time $2T$,  to make a matter wave interferometer that is inertially sensitive~\citep{PhysRevLett.67.181}. 

To measure horizontal acceleration, the atoms fall along a path perpendicular [transverse] to the optical beam used to drive the stimulated Raman transitions. For an interferometer time $T$ that is typically $10$'s of milliseconds, the cloud will fall by a few centimeters. If the local magnetic field varies between the atom positions at each pulse, this will result in a small change in the internal energy levels of the atoms via the second-order Zeeman shift [$575$\,Hz/G$^2$ for the clock transition in $~^{87}$Rb~\citep{Steck}]. This imprints an additional phase on the atomic wavepackets that mimics an inertial signal. These magnetic field gradients, with respect to the trajectory of the atoms during interferometry, represent a significant potential source of uncertainty for quantum inertial sensors, particularly when deployed in a setting with dynamic magnetic field variations. 

To support development of sensors that can work outside of the laboratory, in the presence of dynamic magnetic fields and gradients, we have designed coils to null uniform and linear gradient magnetic fields directed along the $z$-coordinate. We define the fields we wish to null in terms of spherical harmonics [see \cite{10154546} for an introduction to spherical harmonic decomposition in coil design]. They are
\begin{align}
\label{eq.target1} \mathbf{B}_{1,0}&=B_{1}\mathbf{\hat{z}}, \\ 
\label{eq.target2} \mathbf{B}_{2,-1}&=B_{2}\left( z\mathbf{\hat{y}} + y\mathbf{\hat{z}}\right),
\end{align}
where $B_1$ and $B_2$ are scalings and $\left(\mathbf{\hat{x}},\mathbf{\hat{y}},\mathbf{\hat{z}}\right)$ are Cartesian unit vectors. The $\mathbf{B}_{2,-1}$ spherical harmonic contains both a $\mathrm{d}B_z/\mathrm{d}y$ variation and a $\mathrm{d}B_y/\mathrm{d}z$ variation to satisfy Gauss' law for magnetism, $\nabla\cdot\mathbf{B}=0$. This additional gradient does not currently limit  sensor performance.

\subsection{Mathematical model}
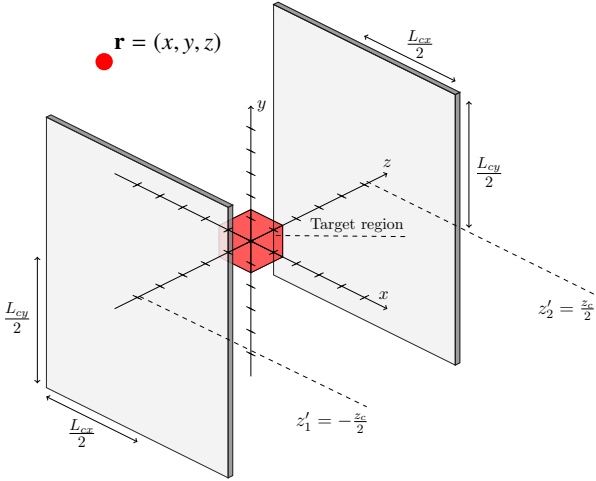
\begin{figure}[h!]
	\centering
    \includegraphics[width=0.9\columnwidth]{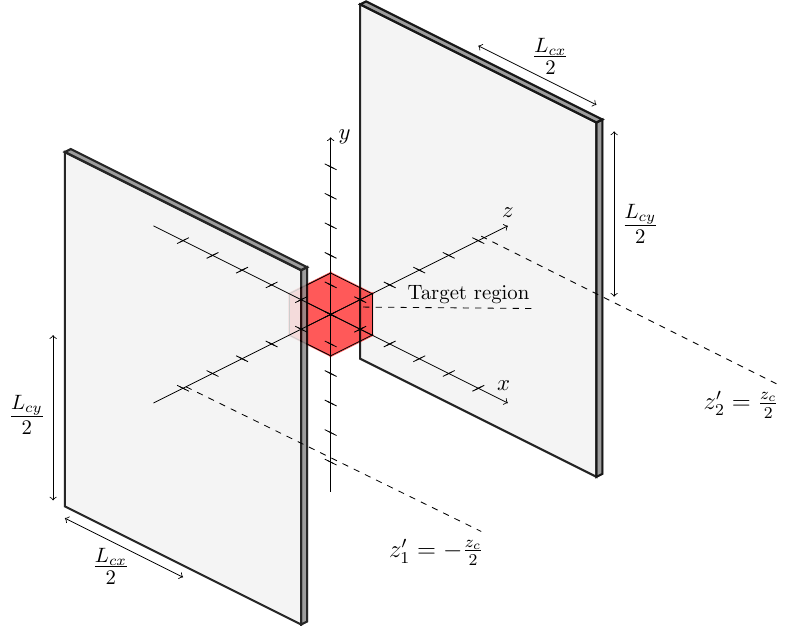}
	% \scalebox{0.7}{\input{tikz/bi_planar}}
	\caption{Rectangular bi-planar coils [grey] of dimension $L_{cx}{\times}L_{cy}$ which are separated by an axial position $z_c$ and centred about the origin. The coils are designed to generate a target magnetic field within the target region [red cube]. Measurement coordinates in free space are denoted by $\mathbf{r} = (x,y,z)$ and coil coordinates are denoted by $\mathbf{r'} = (x',y',z')$.}
	\label{fig.bi_planar}%
	
	% Overlay the 2D TikZ picture (red circle)
	\begin{tikzpicture}[overlay]
		  \draw[thick, red, fill=red, xshift=-130pt, yshift=170pt] (2,2) circle (0.1) node[black, right, font=\small, yshift=7pt]{$\mathbf{r} = (x,y,z$)};
	\end{tikzpicture}
\end{figure}
Let us consider a pair of rectangular current-carrying surfaces on the $xy$-plane with $x$-extent, $L_{cx}$, and $y$-extent, $L_{cy}$. The planes are centred along the $z$-axis at positions $z'_1=-z_c/2$ and $z'_2=+z_c/2$, as shown in Figure \ref{fig.bi_planar}.

Following~\cite{10.1088/0022-3727/19/8/001}, we shall relate the magnetic vector potential, $\mathbf{A}\left(\mathbf{r}\right)$, at a measurement point $\mathbf{r} = (x,y,z)$, to the current density, $\mathbf{J}\left(\mathbf{r}'\right)$, at a point $\mathbf{r'} = (x',y',z')$, through the integral form of Poisson's equation,
\begin{align} \label{eq.AGreen}
    \mathbf{A}\left(\mathbf{r}\right)&=-\frac{\mu_0}{4\pi} \int_{r'} \mathrm{d}^3\mathbf{r'}\ \frac{\mathbf{J}\left(\mathbf{r'}\right)}{\left|\mathbf{r}-\mathbf{r'}\right|} \nonumber \\
    &=\mu_0 \int_{r'} \mathrm{d}^3\mathbf{r'}\ G\left(\mathbf{r},\mathbf{r'}\right) \mathbf{J}\left(\mathbf{r'}\right).
\end{align}
As we wish to design coils bound to the $xy$-plane, we choose a Green's function, $G\left(\mathbf{r},\mathbf{r'}\right)$, which is separable in Cartesian coordinates~\citep{Yoda1990},
\begin{align} \label{eq.AGreen2}
G\left(\mathbf{r},\mathbf{r'}\right) &= \frac{1}{8\pi^3} \iiint_{-\infty}^{\infty}\mathrm{d}\mathbf{k}^{3}  \frac{e^{i(k_x(x-x')+k_y(y-y')+k_z(z-z'))}}{\sqrt{k_x^2+k_y^2+k_z^2}},
\end{align}
where $\mathbf{k} = (k_x,k_y,k_z)$. 

By substituting equation~\eqref{eq.AGreen2} into equation~\eqref{eq.AGreen} and then taking its curl, we determine the magnetic field, $\mathbf{B}\left(\mathbf{r}\right)=\nabla\wedge\mathbf{A}\left(\mathbf{r}\right)$, [where $\wedge$ represents the vector cross product] generated by a single plane at axial position $z'=z_c$. Expanding the magnetic field in Cartesian coordinates, $\mathbf{B}\left(\mathbf{r}\right)=\mathbf{B}_x\left(\mathbf{r}\right)~\mathbf{\hat{x}} + \mathbf{B}_y\left(\mathbf{r}\right)~\mathbf{\hat{y}} + \mathbf{B}_z\left(\mathbf{r}\right)~\mathbf{\hat{z}}$, we find
\begin{multline}\label{eq.Bx.rel.2}
    B_{x}\left(x,y,z\right) = \\ -\frac{\mu_{0}\tilde{z}}{2\pi|\tilde{z}|} \int_{-\infty}^{\infty}\mathrm{d}k_x \ \int_{-\infty}^{\infty}\mathrm{d}k_y \  \tilde{J}_{y}\left(k_x,k_y\right) e^{i(k_x x + k_y y)} 
    e^{-\sqrt{k_x^2+k_y^2}|\tilde{z}|},
\end{multline}
\begin{multline}\label{eq.By.rel.2}
    B_{y}\left(x,y,z\right) = \\ \frac{\mu_{0}\tilde{z}}{2\pi|\tilde{z}|} \int_{-\infty}^{\infty}\mathrm{d}k_x \ \int_{-\infty}^{\infty}\mathrm{d}k_y \  \tilde{J}_{x}\left(k_x,k_y\right) e^{i(k_x x + k_y y)} 
    e^{-\sqrt{k_x^2+k_y^2}|\tilde{z}|},
\end{multline}
\begin{multline}\label{eq.Bz.rel.1}
    B_{z}\left(x,y,z\right) = \frac{i\mu_{0}}{2\pi} \int_{-\infty}^{\infty}\mathrm{d}k_x \ \int_{-\infty}^{\infty}\mathrm{d}k_y \ \\
    \left(k_x \tilde{J}_{y}\left(k_x,k_y\right) - k_y \tilde{J}_{x}\left(k_x,k_y\right)\right) e^{i(k_x x + k_y y)}
    \frac{e^{-\sqrt{k_x^2+k_y^2}|\tilde{z}|}}{\sqrt{k_x^2+k_y^2}},
\end{multline}
where $\tilde{z} = z - z_c$ is the axial distance from the coil plane to the measurement point. The magnetic field is determined by the Fourier transform of the current density on the plane, $\mathbf{J}\left(\mathbf{r}'\right)=\mathbf{J}_x\left(\mathbf{r}'\right)~\mathbf{\hat{x}} + \mathbf{J}_y\left(\mathbf{r}'\right)~\mathbf{\hat{y}}$, where $\mathbf{J}_{x/y}\left(\mathbf{r}'\right)$ are the $x/y$ components. The Fourier transforms are defined by
\begin{multline}\label{eq.FT}
        \tilde{J}_{x/y}\left(k_x,k_y\right) = \\ \frac{1}{4\pi^2}\int_{-\infty}^{\infty}\mathrm{d}x' \ \int_{-\infty}^{\infty}\mathrm{d}y' \ e^{-i\left({k_x}x'+{k_y}y'\right)}J_{x/y}\left(x',y'\right).
\end{multline}

As current is contained on the planar surface, $\nabla\cdot\mathbf{J}\left(\mathbf{r}'\right)=0$, the current density may be defined in terms of the normal projection of the gradient of a single valued streamfunction, $\nabla\varphi\left(x',y'\right) \wedge  \mathbf{\hat{z}}=0$, where
\begin{equation}\label{eq.jsrect}
    J_{x}\left(x',y'\right) = \frac{\partial\varphi\left(x',y'\right)}{\partial y'}, \qquad J_{y}\left(x',y'\right) = -\frac{\partial\varphi\left(x',y'\right)}{\partial x'}.
\end{equation}
Here, we choose to represent the streamfunction as a set of Fourier modes which satisfy equation~\eqref{eq.jsrect} to bound the current to the plane,
\begin{multline}\label{eq.streamyrect}
    \varphi\left(x',y'\right) = \left[\mathcal{H}\left(x'-\frac{L_{cx}}{2}\right)-\mathcal{H}\left(x'+\frac{L_{cx}}{2}\right)\right] \\ 
    \left[\mathcal{H}\left(y'-\frac{L_{cy}}{2}\right)-\mathcal{H}\left(y'+\frac{L_{cy}}{2}\right)\right] \times \frac{\sqrt{L_{cx}L_{cy}}}{\pi} \\
    \sum_{n_x=1}^{N_x}\sum_{n_y=1}^{N_y}\ A_{{n_x}{n_y}} \sin\left(\frac{n_x\pi\left(x'-L_{cx}/2\right)}{L_{cx}}\right)\sin\left(\frac{n_y\pi\left(y'-L_{cy}/2\right)}{L_{cy}}\right).
\end{multline}
The streamfunction contains modes that oscillate along $x'$ and $y'$ up to orders $N_x$ and $N_y$, respectively. The symbol $\mathcal{H}(x)$ represents a Heaviside step function and $A_{{n_x}{n_y}}$ are the weightings of each Fourier mode, which we shall refer to as \emph{Fourier coefficients}. In \ref{App.maths}, we calculate the Fourier transforms of the current densities expressed in terms of the Fourier modes. Then, we substitute these into equations~\eqref{eq.Bx.rel.2}--\eqref{eq.Bz.rel.1} and rearrange terms. The resulting master equations, \eqref{Bx_pseudo_final}--\eqref{Bz_pseudo_final}, encode the magnetic field generated by each Fourier mode, bound to a pair of rectangular coils [Figure~\ref{fig.bi_planar}] either carrying the same or opposite current flow, in terms of Fourier space integrals.

\subsection{Bi-planar coil optimisation}
We use this mathematical model to design bi-planar coils to generate specific target fields by finding sets of Fourier coefficients that generate optimal current flows. Any target field can be constructed from a superposition of solutions where the planar pair carry either equal or opposite current. If the target field is simple and the region where the target field is placed centrally and symmetrically between the coils, we can exploit symmetries in the axial coordinate to vastly reduce the number of modes which are computed. This maximises computational efficiency and enables more relevant modes to be included, thereby maximising the quality of the magnetic field generated.

The uniform axial field, $\eqref{eq.target1}$, between the planes is symmetric along the $z$-axis and therefore requires that the coil planes carry equal currents. The $x'$ and $y'$ varying modes in the streamfunction on the coil planes, \eqref{eq.streamyrect}, must also be symmetric so that the $B_x$ and $B_y$ components generated by opposite sides of each plane cancel as the current flow is determined by the streamfunction gradient across the coil, \eqref{eq.jsrect}. The streamfunction, \eqref{eq.streamyrect}, is intentionally constructed such that modes, with orders where if $n_x$ or $n_y$ is odd, are symmetric across the surface with respect to $x'$ or $y'$, respectively. Therefore, the uniform $B_z$ coil only requires modes to be optimised where both $n_x$ and $n_y$ are odd. 

On the other hand, to generate the linear gradient in the axial field with respect to transverse position, \eqref{eq.target2}, the streamfunction must be symmetric along the $x'$ coordinate and anti-symmetric along the $y'$ coordinate to generate the $B_y$ component, and so modes are optimised only where $n_x$ is odd and $n_y$ is even. The coil planes must carry equal currents to generate this gradient linearly along $z$.

Once we have determined which modes to include, we follow the standard approach~\citep{Holmes2019,PhysRevApplied.15.054004} and employ least-squares minimisation~\citep{boggs_tolle_1995} of the squared deviation between the target field and the field generated by each relevant mode within the target region to determine the optimal values of the Fourier coefficients, $A_{{n_x}{n_y}}$. In order to manufacture the coils using standard techniques, we regularise the least-squares optimisation by employing a regularisation parameter related to the curvature of the streamfunction~\citep{10.1088/0022-3727/35/9/303},
\begin{equation} \label{eq.crv}
    C=\gamma\int_{r'}\mathrm{d}^2\mathbf{r}'\ |\mathbf{\nabla}\varphi(\mathbf{r}')|^2,
\end{equation}
where $\gamma$ is a tuneable parameter to control the strength of the regularisation. In ~\ref{App.maths}, we integrate over the surface of the planes to find
\begin{align}\label{curvature}
    C &= \frac{\gamma\pi^2}{4} \sum_{n_{x}=1}^{N_{x}}\sum_{n_{y}=1}^{N_{y}}\ A_{n_{x}{n_{y}}}^2\left(n_{x}^2\frac{L_{cy}}{L_{cx}} + n_{y}^2\frac{L_{cx}}{L_{cy}}\right)^2.
\end{align}
Once the Fourier coefficients have been calculated, they are used to calculate the planar, single-valued streamfunction in equation~\eqref{eq.streamyrect} at both planar positions. The streamfunction is then contoured to generate discrete wire patterns which represent the path of current on the coil's surface~\citep{PhysRevApplied.15.054004}.

\section{Materials \& Methods}
\subsection{Design}
\begin{figure}[hbt!]
\centering
\includegraphics[width=\columnwidth]{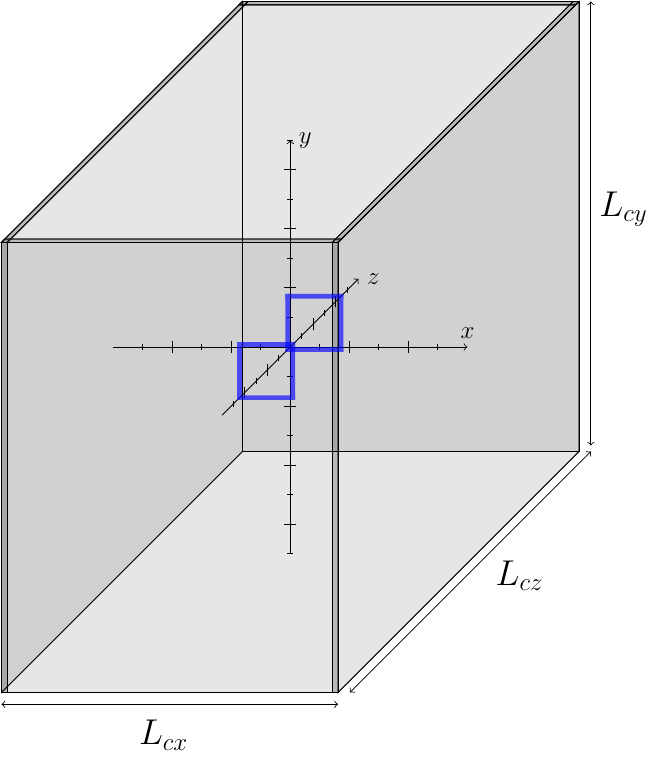}
% \scalebox{0.6}{\input{tikz/imperial_coords}}
\caption{Four panel system for generating a uniform $B_{z}$ field and a linear $\mathrm{d}B_{z}/\mathrm{d}y$ field gradient. The coils are mounted on the quantum sensor mounting frame [grey] of volume $L_{cx} \times L_{cy} \times L_{cz}$, where $\left(L_{cx}, L_{cy}, L_{cz}\right)=\left(570, 762, 1060\right)$~mm. The panels are all situated within the $xy$-plane, with two each at $z={\pm}L_{cz}/2$. The previous $B_z$ coil system of side lengths $180 \times 184$ mm\textsuperscript{2} and separation $424$~mm is also displayed [blue outline].}
\label{fig.Imperial_rig}
\end{figure}
We now apply the mathematical framework in section~\ref{sec:theory} to design two bi-planar coil systems to independently and simultaneously generate a uniform axial field, \eqref{eq.target1}, and a linear axial field gradient with respect to transverse position, \eqref{eq.target2}. The coil system is designed to stabilise these variations throughout the whole atomic path, which aligns along the transverse $y$-coordinate and is contained within a central volume of $60 \times 60 \times 60$~mm\textsuperscript{3}. The atomic path is located centrally inside the experimental housing and so the target region is equidistant between the panels. The required strengths of the target magnetic fields are determined by measured values during a previous field test of the inertial sensor. The generated uniform axial field must be stronger than Earth's magnetic field, $B_1 > 50$~$\mu$T, and the gradient stronger than $B_2>50~\mu$T/Am. This must be achieved using applied currents $I\leq5~\mathrm{A}$ due to limits to the quantum sensors' on-board current drive system. The uniform axial field is required to deviate by $<4$\% within the target region to out-perform the previous on-sensor compensation coil which was based on a rectangular pair carrying equal currents, as displayed in Figure~\ref{fig.Imperial_rig}. This coil is closer to the target region than the bi-planar coil set because of the large power-efficiency requirement; the bi-planar continuum coil optimisation allows current to be distributed across the surface so the panels can be placed much further away.

The coil designs are generated by first solving the master equations, \eqref{Bx_pseudo_final}--\eqref{Bz_pseudo_final}, using numerical integration with the \texttt{integral2()} function in \textit{MATLAB R2023a} up to mode order $N_x=N_y=10$. The infinite limits of $k_x$ and $k_y$ in the numerical integration are approximated as $\left(k_x,k_y\right)\rightarrow10^3$ with convergence errors less than one part in $10^6$ within the target region. The least-squares optimisation is implemented using simple matrix manipulation and the optimal Fourier coefficients are then substituted into the streamfunction, \eqref{eq.streamyrect}, which is sampled densely across the coil's surface. The optimal streamfunction is discretised into contour levels using the \textit{MATLAB R2023a} \texttt{contour()} function. The number of contour levels is selected to maximise the power efficiency of the coils, while keeping the total wire length on each panel less than $160$~m for ease-of-manufacture. The final wire patterns consist of the contoured streamlines of the streamfunction connected in series at their closest point, and then traced with a return wire to cancel unwanted fields generated by the connections.

\begin{figure}[t]
    \centering
    \includegraphics[width=0.48\textwidth]{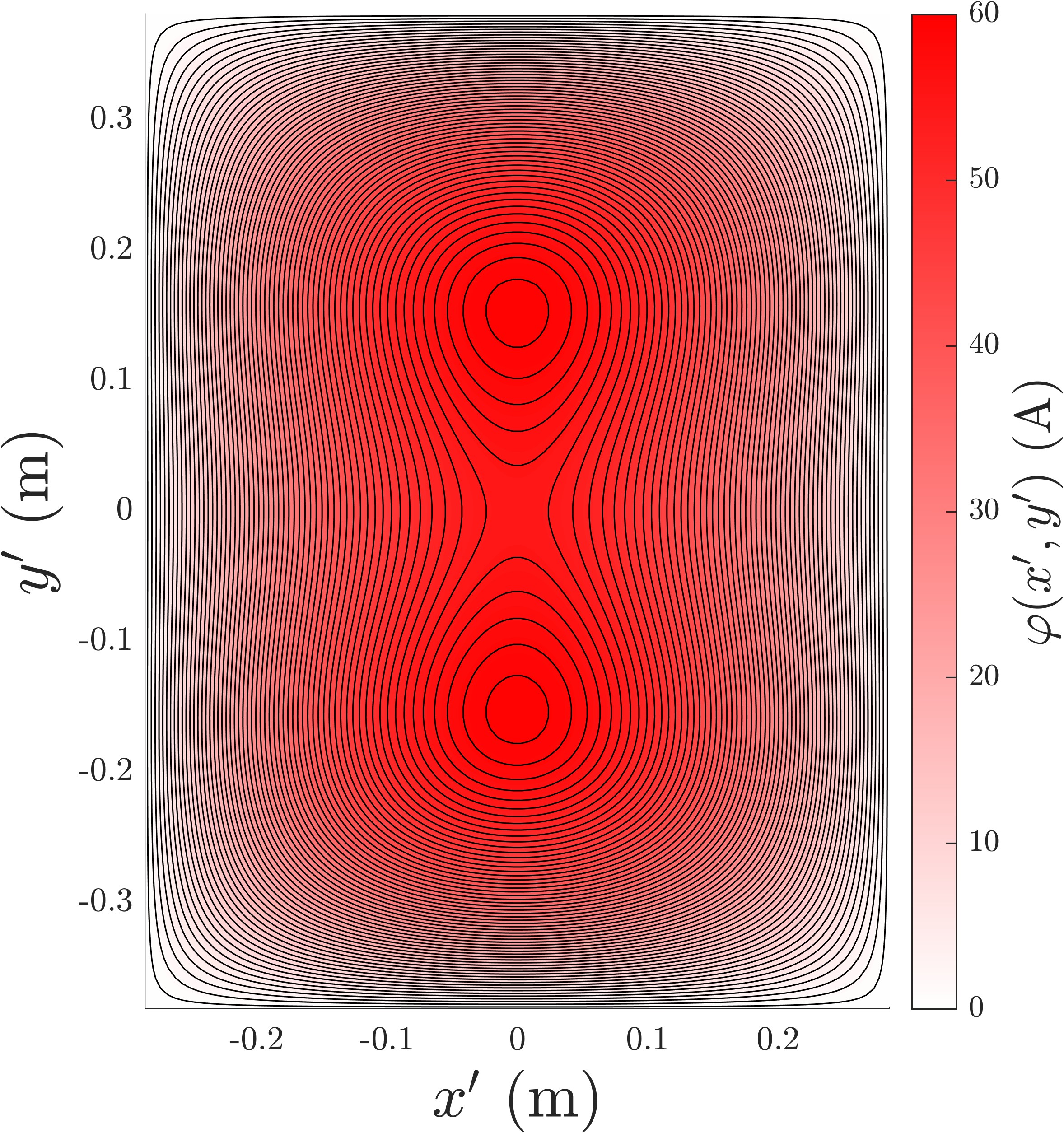}
    \caption{Wire layouts in which current flows on a panel of dimensions $P_{xy} =  570$ x $762$ mm$^{2}$ [black solid and dashed lines represent opposite directions of current flow] as determined by the streamfunction [red positive with intensity of colour proportional to the magnitude]. The wire layouts are optimised to generate a uniform field, $B_{z}$.}
    \label{fig:Bz Stream}
\end{figure}
\begin{figure}[t]
    \centering
    \includegraphics[width=0.48\textwidth]{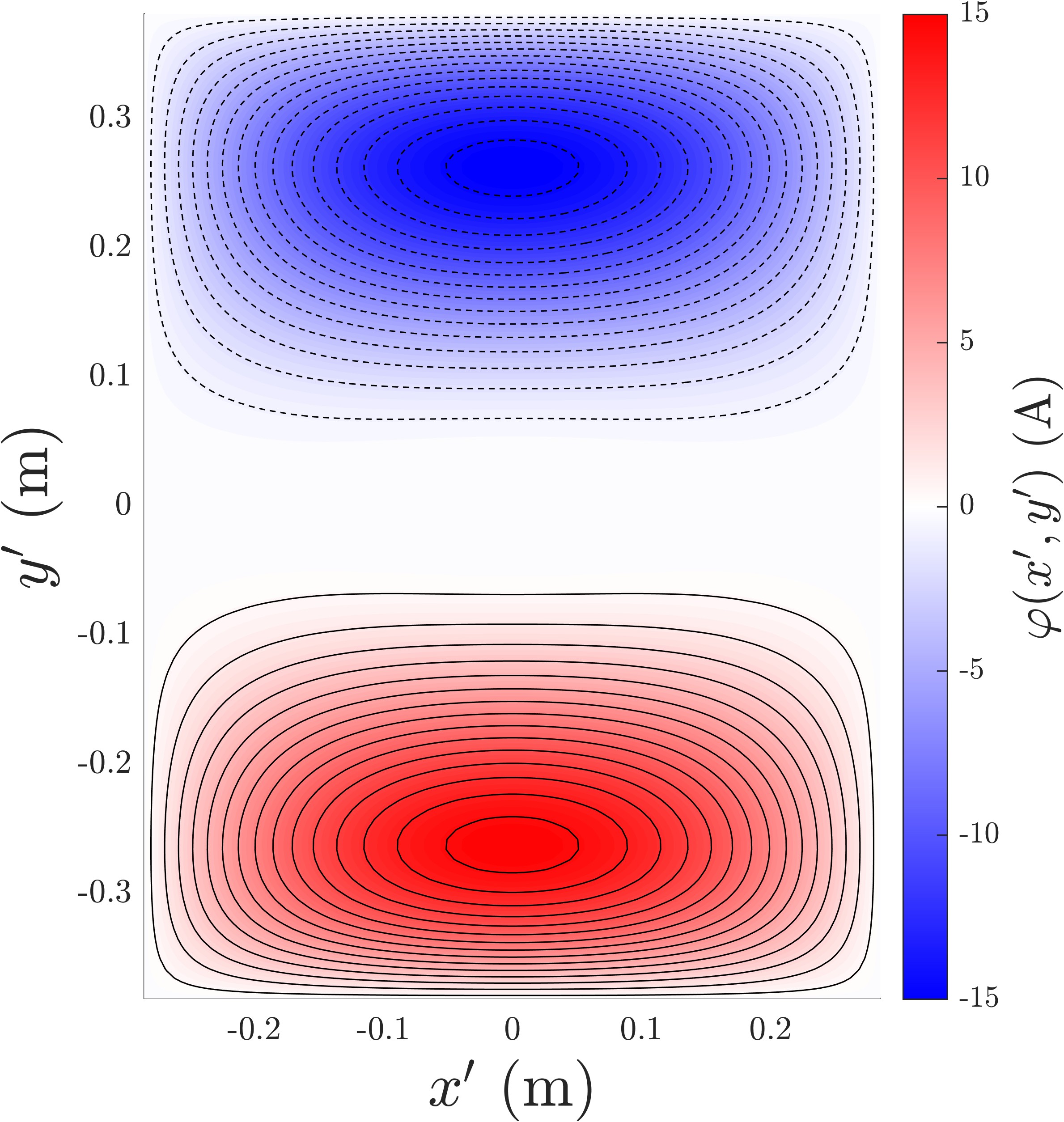}
    \caption{Wire layouts in which current flows on a panel of dimensions $P_{xy} =  570$ x $762$ mm$^{2}$ [black solid and dashed lines represent opposite directions of current flow] as determined by the streamfunction [red positive and blue negative with intensity of colour proportional to the magnitude]. The wire layouts are optimised to generate a gradient field, $\mathrm{d}B_z/\mathrm{d}y$.}
    \label{fig:dBy Stream}
\end{figure}
The streamfunctions and discrete coil designs to generate the $\mathbf{B}_{1,0}$ and $\mathbf{B}_{2,-1}$ target fields are presented in Figures~\ref{fig:Bz Stream}~and~\ref{fig:dBy Stream}, respectively. As detailed in Figure~\ref{fig.Imperial_rig}, both coils are housed on the $xy$-plane on panels of side lengths $L_{cx}=570$~mm and $L_{cy}=762$~mm inside the mounting frame of axial length $L_{cz}=1060$~mm. Each coil contains two separate panels housed at positions $z_c=\pm L_{cz}/2$ within the mounting frame. While in principle the designs could be shared on the same panel, the wires would deviate from the ideal pattern at the points where the wires cross. There are a large number of these crossings because the streamfunction is contoured to a high density due to the high power-efficiency required to meet the specification. Therefore, we choose to make the designs on separate panels. Notably, the $yz$-planar faces of the mounting structure are free of any optical access constraints, and so could be used to house further coil systems in future iterations of the magnetic field stabilisation system.

\subsection{Manufacture}
\begin{figure}[ht!]
\begin{center}
         \includegraphics[width=0.42\textwidth]{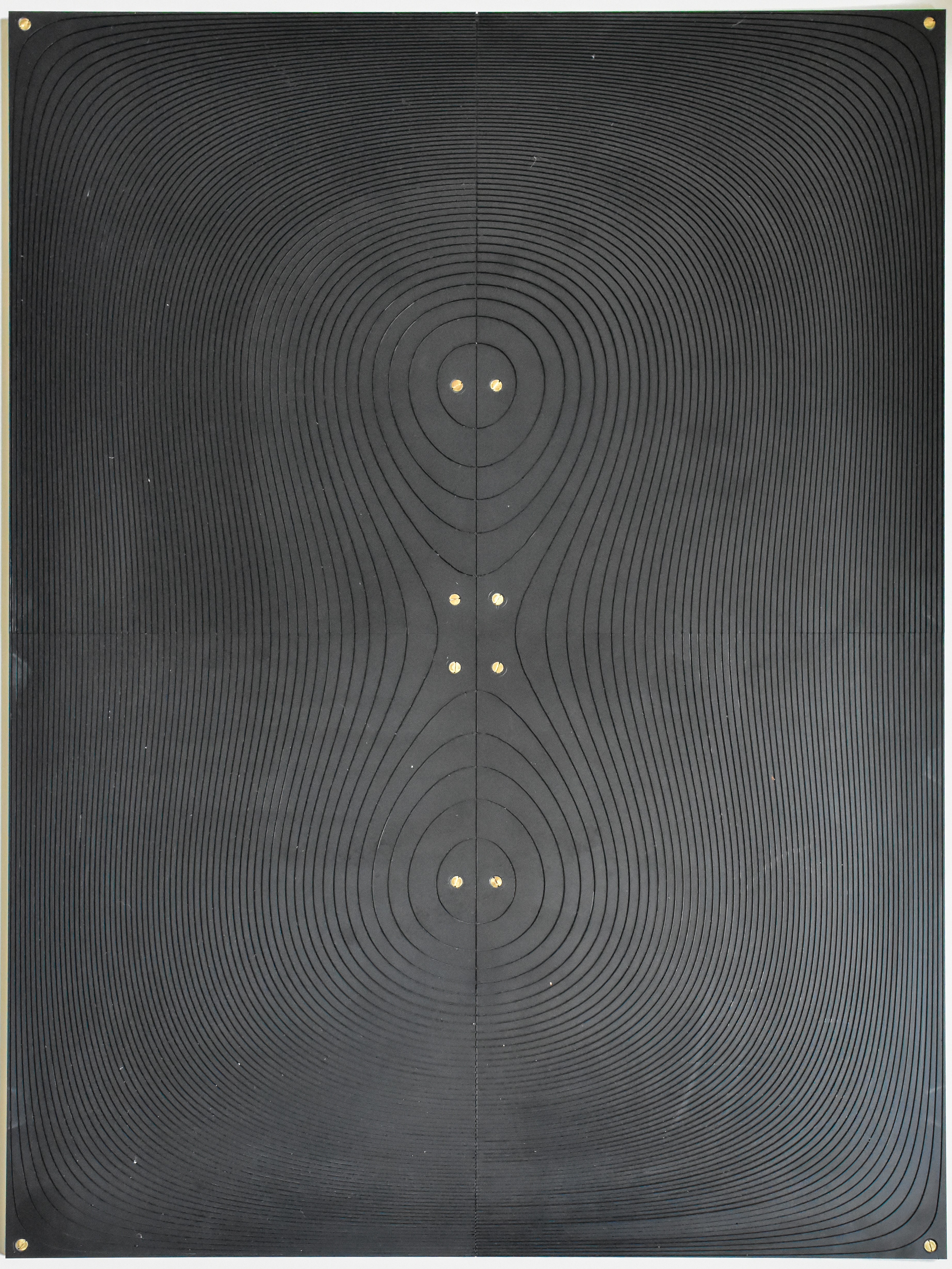} \\ (a) \\
         \includegraphics[width=0.42\textwidth]{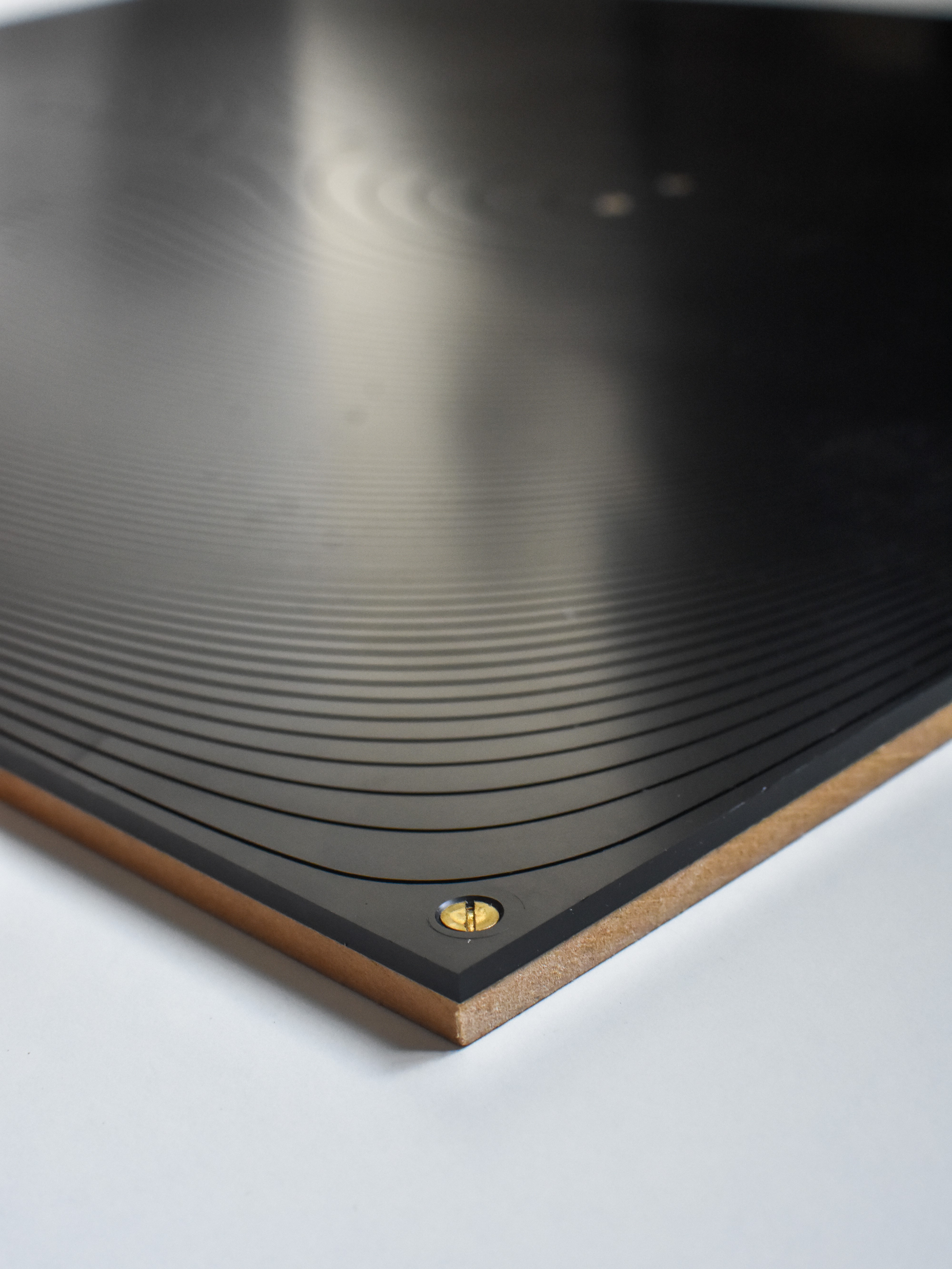} \\ (b)
\caption{Manufacture of the $\mathbf{B}_{1,0}$ coil: (a) Overview of four acrylic sections aligned together to form one panel before winding. (b) The acrylic is mounted onto a backing fibreboard of $6~\mathrm{mm}$ thickness with brass screws.}
\label{fig.Bz_manufacture}
\end{center}
\end{figure}
The wire coordinates are imported into \textit{SOLIDWORKS} Computer-Aided Design (CAD) software alongside a 3D panel representing the coil surface. The \textit{SOLIDWORKS} \texttt{cut-sweep()} function is then used to etch the wire patterns as grooves in the 3D panel of width $0.85~\mathrm{mm}$. The connections between the contours are encoded at double the depth to allow the return wire to pass underneath to mitigate the erroneous field. The CAD designs are manufactured by Computer Numerical Control (CNC) routing of acrylic sheets using a \textit{Haas Automation Inc.} TM-2CE Toolroom Milling Machine. As presented in Figure~\ref{fig.Bz_manufacture}, each panel was made of four sheets screwed together due to limits to the CNC machine bed size. Once routed, the panels are wound with 22 AWG [$0.63~\mathrm{mm}$ diameter] enamelled copper wire which passes a maximum current of $I=6$~A. The wires are secured in the grooves using a hot glue gun and superglue. The planes are attached to the mounting frame using brass screws.

\subsection{Characterisation}
We use a \textit{Bartington Instruments} Mag-13 MCL-100 low noise three-axis magnetometer [dynamic range $\pm 100~ \mu\mathrm{T}$, noise floor $<10$~pT/$\sqrt{}$Hz] to measure the magnetic fields generated by the coils. The magnetometer is mounted onto three \textit{ThorLabs} motorised translation stages to allow the magnetometer to move between the coils. Custom aluminium connectors are used to attach the stages to one another to allow the magnetometer to move through a $50\times300\times300$~mm\textsuperscript{3} volume of space without needing to re-align the stages. The stages are then mounted onto an optics board suspended between the coils using optical dovetail railings. This is mounted directly to the skeleton rig which also includes the coil panels. This helps reduce vibrations that would cause magnetometer and coil panels to move relative to each other and introduce measurement error.

During characterisation, we measure the magnetic field generated by both coil sets under an applied sinusoidal current of amplitude $500~\mathrm{mA}$ and frequency $6~\mathrm{Hz}$. The current is monitored by interrogating the voltage across a $100$~m$\Omega$ resistor connected in series to each coil using a \textit{National Instruments} NI-USB 6212 Data Acquisition system (DAQ), which is also used to simultaneously process data from the magnetometer. Measurements along the $y$-axis are performed at $1~\mathrm{mm}$ increments and data is sampled at $f=1$~kHz for $20~\mathrm{s}$ per position. Alongside the axis sweeps, we measure the magnetic field inside $zy$-planes crossing the centre of the target region in $10~\mathrm{mm}$ increments. The planes are positioned at the centre of the target region, $x=0~\mathrm{mm}$, and its edge, $x = 30~\mathrm{mm}$. During data processing, we calculate the Fourier transform of the measured data to determine the magnitude of the target field generated by the coil per unit current at the $6$~Hz drive frequency. This alleviates the need to calculate measurement offsets.

\section{Results}
\begin{table*}[hbt]
    \centering
    \begin{tabular}{|c|c|c|c|}
    \hline
        \multicolumn{2}{|c|}{\textbf{Field profile}} &  $B_1$ ($\mu$T/A) & $B_2$ ($\mu$T/Am)  \\
    \hline \hline
        \multirow{2}{*}{Theory} & $\mathbf{B}_{1,0}$ & $22.81\pm0.01$~ & $0$  \\   
    \cline{2-4}
        & $\mathbf{B}_{2,-1}$ & $0$ & $11.0\pm0.01$  \\  
    \hline
        \multirow{2}{*}{Experimental} & $\mathbf{B}_{1,0}$ & $22.55\pm0.02$~ & $\left(0 \pm1\right) \times 10^{-3}$  \\   
    \cline{2-4}
        & $\mathbf{B}_{2,-1}$ & $-\left(7\pm1\right) \times 10^{-3}$ & $10.6\pm0.1$  \\  
    \hline
    \end{tabular}
    \caption{Fitting parameters of $B_z\left(y\right)=B_1 + B_2 y$ along the $y$-axis of the target region from theoretical analytical results calculated using equation~\eqref{Bz_pseudo_final} and experimental data. Errors are calculated with $\pm 2\sigma$ confidence.}
    \label{tab:lineoffit}
\end{table*}
\begin{figure}[h]
    \centering
    \includegraphics[width=0.48\textwidth]{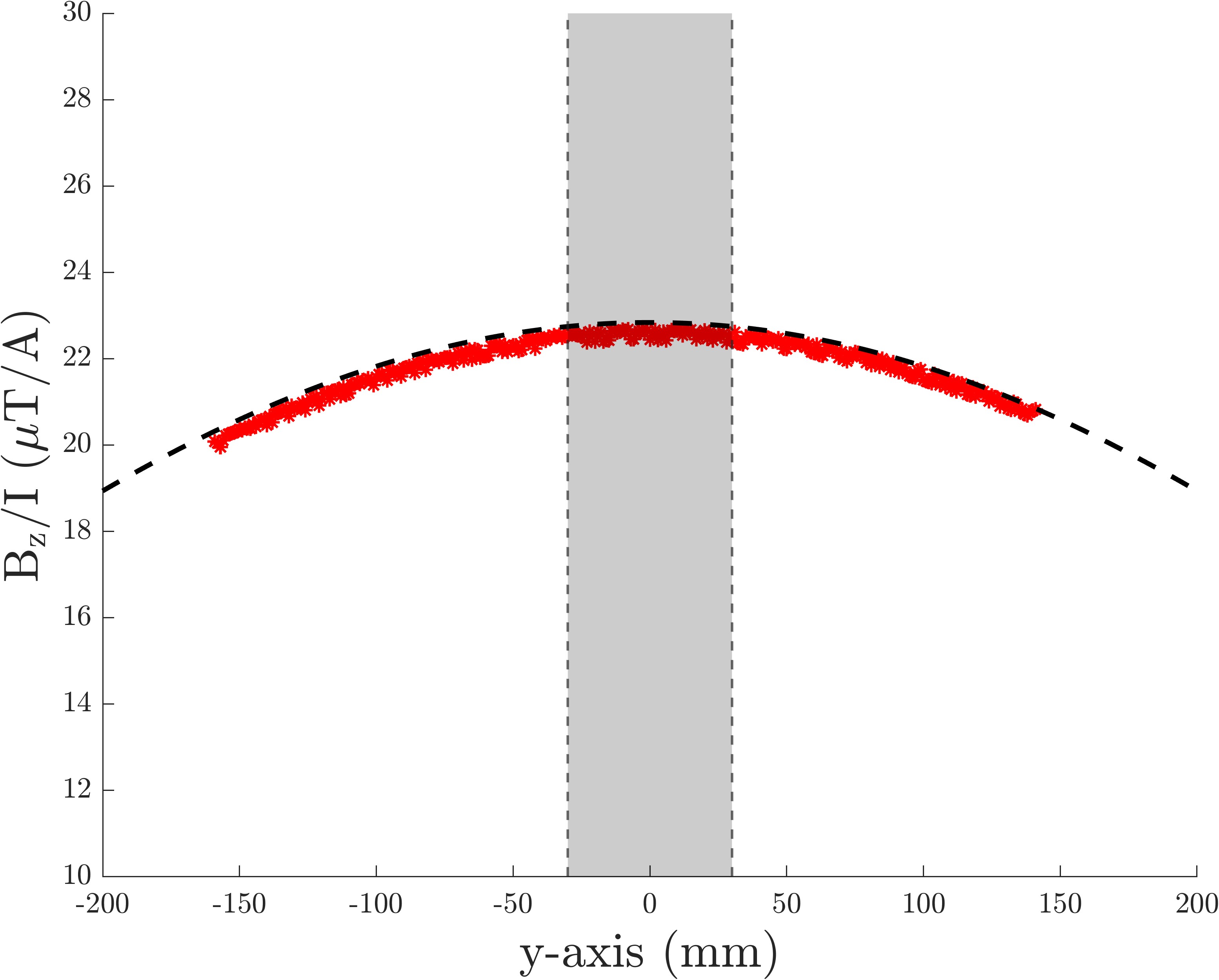}
    \caption{$B_z/\mathrm{I}$ field profile at $x = 0~\mathrm{mm}$ and $z = 0~\mathrm{mm}$ along the $y$-axis for the $\mathbf{B}_{1,0}$ coil. Red shows experimental data and black dashed shows analytical results calculated using equation~\eqref{Bz_pseudo_final}. Target region is enclosed with grey fill.}
    \label{fig:Bz field profile}
\end{figure}
\begin{figure}[h]
    \centering
    \includegraphics[width=0.48\textwidth]{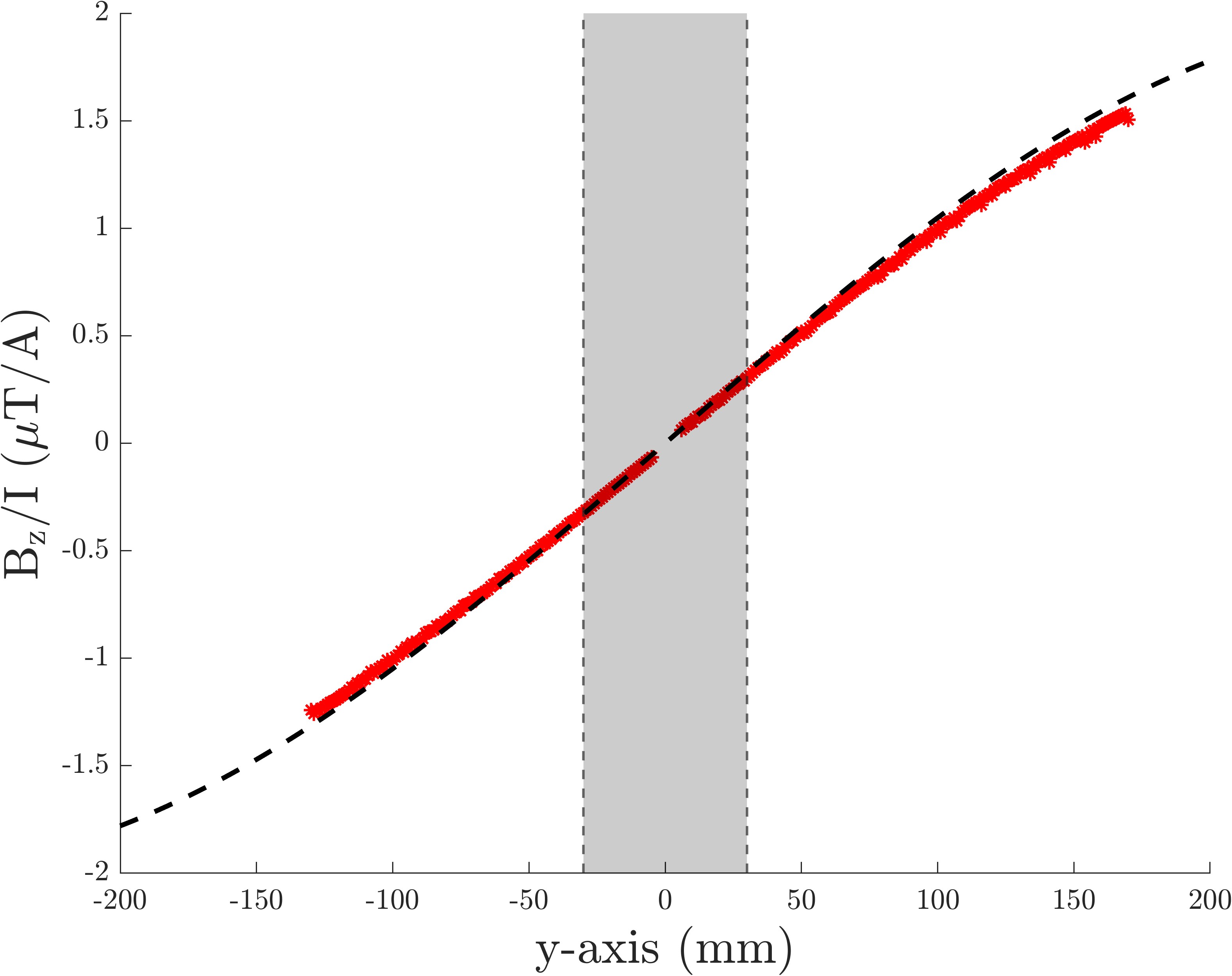}
    \caption{$B_z/\mathrm{I}$ field profile at $x = 0~\mathrm{mm}$ and $z = 0~\mathrm{mm}$ along the $y$-axis for the $\mathbf{B}_{2,-1}$ gradient coil. Red shows experimental data and black dashed shows analytical results calculated using equation~\eqref{Bz_pseudo_final}. Target region is enclosed with grey fill.}
    \label{fig:dBz/dy field profile}
\end{figure}
The experimentally measured $B_z$ magnetic field along the $y$-axis is compared to the values predicted by the analytical model for the $\mathbf{B}_{1,0}$ and $\mathbf{B}_{2,-1}$ coils in Figures~\ref{fig:Bz field profile}~and~\ref{fig:dBz/dy field profile}, respectively. From visual inspection, the experimentally measured data show good agreement to the target profiles, although there are offsets between the analytical theory and experimental data. To examine this further, in Table~\ref{tab:lineoffit} we present the parameters for a linear fit to $B_z\left(y\right)=B_1 + B_2 y$ within the target region along the $y$-axis with $\pm 2\sigma$ confidence. As expected, the fits to the experimental data show that the desired harmonics are generated with very low error: $B_1=\left(22.55\pm0.01\right)$~$\mu$T/A for the $\mathbf{B}_{1,0}$ coil and $B_2=\left(10.6\pm0.1\right)$~$\mu$T/Am for the $\mathbf{B}_{2,-1}$ coil. The power efficiency of the coils is greater than the required specification, which strengths of $B_1=112.8$~$\mu$T and $B_2=53$~$\mu$T/m possible with $5$~A of drive current. The offset fields are also very low, e.g. the $B_1=-\left(7\pm1\right)$~nT/A offset of the $\mathbf{B}_{2,-1}$ coil is more than three orders of magnitude smaller than the target $B_1=\left(22.55\pm0.01\right)$~$\mu$T/A magnitude generated by the $\mathbf{B}_{1,0}$ coil.

However, for both the $\mathbf{B}_{1,0}$ and $\mathbf{B}_{2,-1}$ coils, the target fields measured experimentally are weaker than those predicted theoretically. Specifically, the $B_1$ and $B_2$ target fields are respectively $1.1$\% and $3.6$\% smaller in magnitude for the experimental data compared to the theoretical model. This is likely due to challenges with the positional and rotational alignment of the coils around the mounting structure, as well as bowing of the mounting rig due to the weight of the coils. Despite these offsets, the target harmonics are generated with good spatial agreement in the target region because the curvature regularisation, \eqref{curvature}, is strongly enforced during the design process and the target region is far from the coils. This makes the harmonic generation less impacted by small misalignments.

\begin{figure}[b!]
    \centering
    \includegraphics[width=0.48\textwidth]{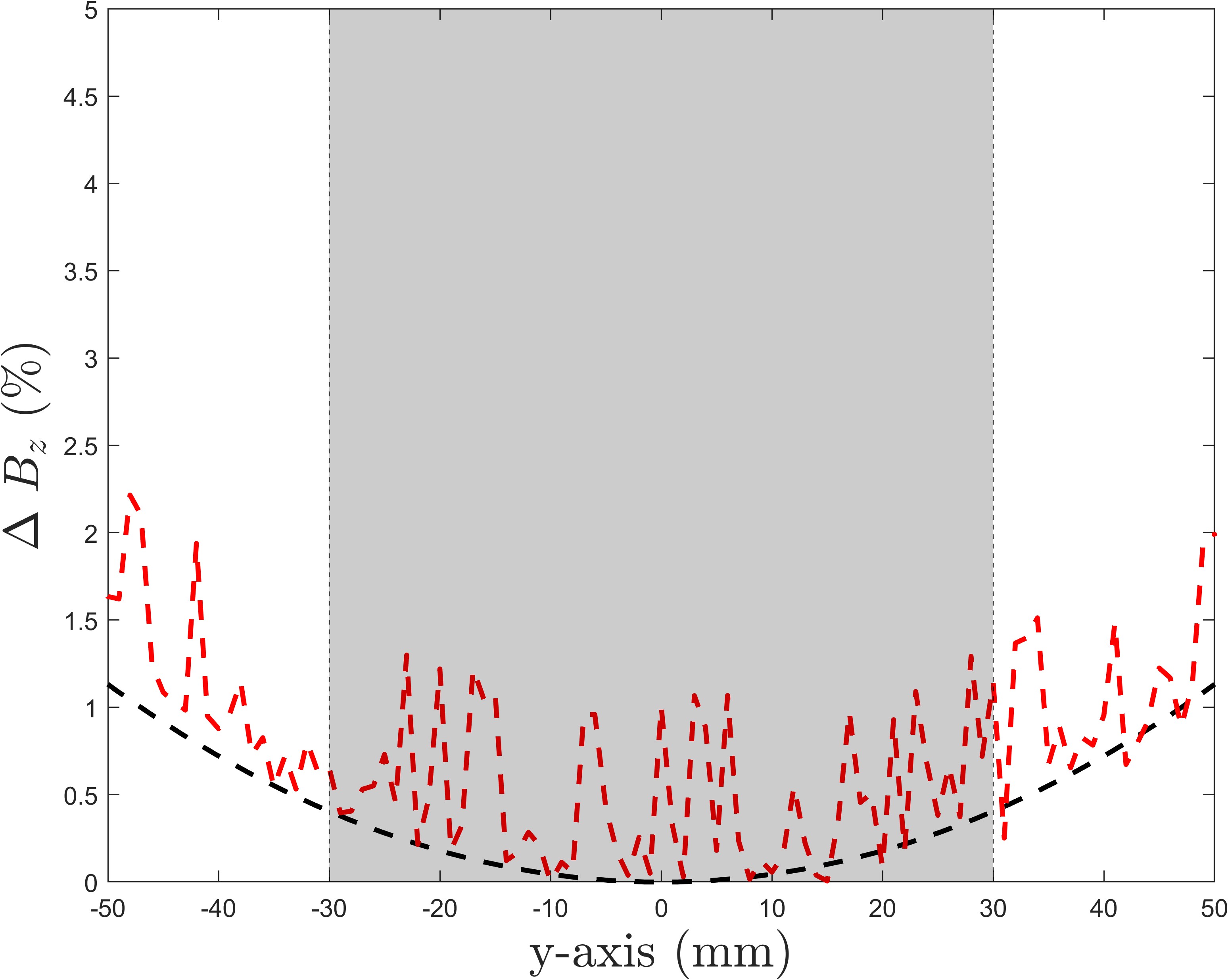}
    \caption{$B_z$ deviation at $x = 0~\mathrm{mm}$ and $z = 0~\mathrm{mm}$ along the $y$-axis for the $\mathbf{B}_{1,0}$ coil. Red shows experimental data and black dashed shows analytical results calculated using equation~\eqref{Bz_pseudo_final}. Target region is enclosed with grey fill.}
    \label{fig:Bz_deviation}
\end{figure}
In Figure~\ref{fig:Bz_deviation}, we present the deviation of the target field from its value at the coil's centre for the theoretical and experimental $\mathbf{B}_{1,0}$ data. The shape of the error profile is very similar for both the theory and experiment. In the theoretical calculations, the maximum deviation of the $B_{z}$ field from the desired profile within the target region is $\mathrm{max}\left(\Delta B_{z}\right)=0.44$\%, while the experimental profile exhibits a higher maximum deviation, $\mathrm{max}\left(\Delta B_{z}\right)=1.30$\%. Examining the deviation in Figure~\ref{fig:Bz_deviation}, we note that the point-wise error is spatially uncorrelated and so is likely due to experimental noise. This is likely because the data measurements were performed in an unshielded environment. Irrespective of this, the measurement maximum field error, $\mathrm{max}\left(\Delta B_{z}\right)=1.30$\%, is significantly lower than the $<4$\% requirement necessary to out-perform the previous onboard rectangular coil set.

\begin{figure}[hbt!]
\begin{center}
  \begin{tabular}{c c} %% tabular useful for creating an array of images 
         \includegraphics[width=0.21\textwidth]{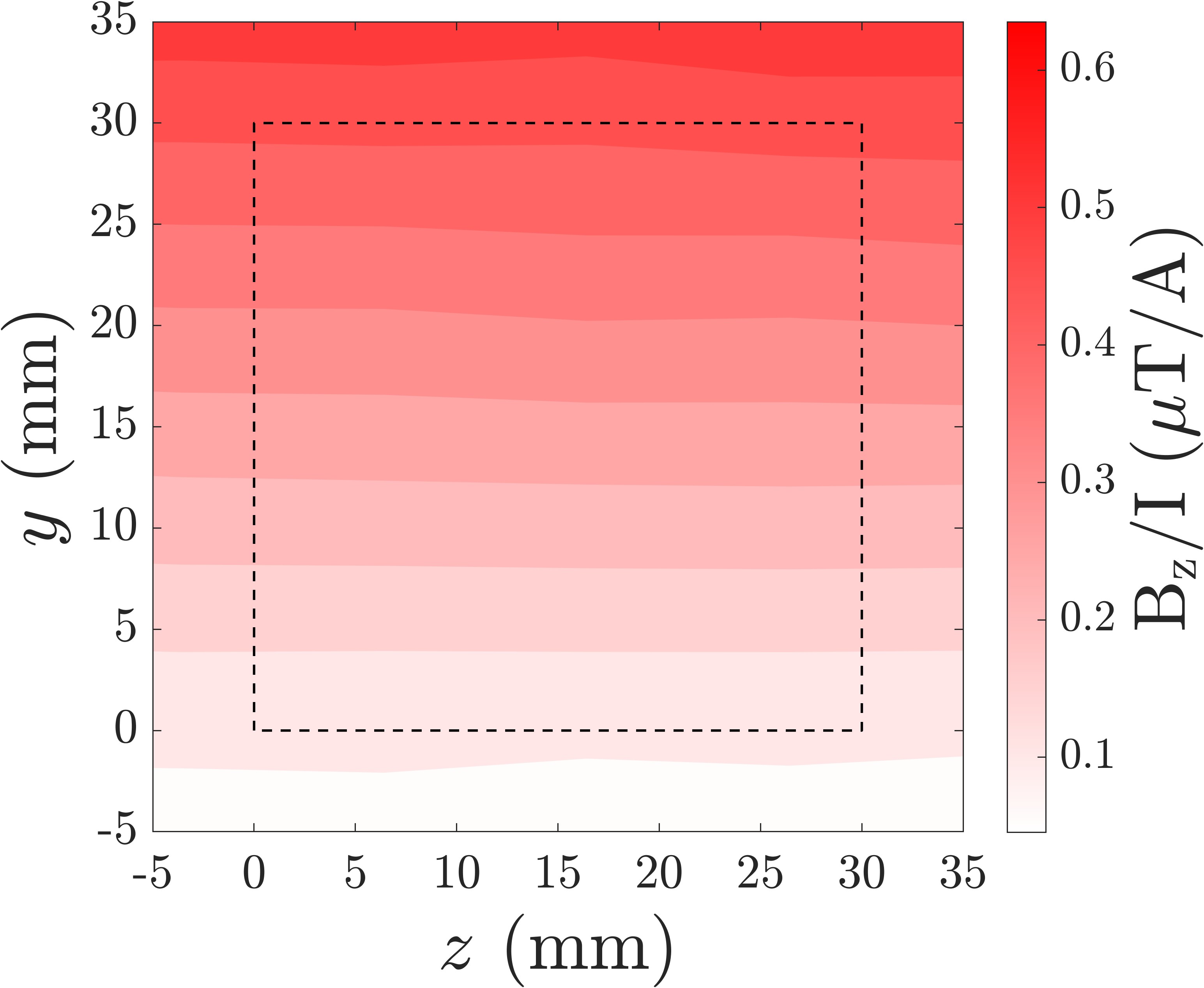} &
         \includegraphics[width=0.21\textwidth]{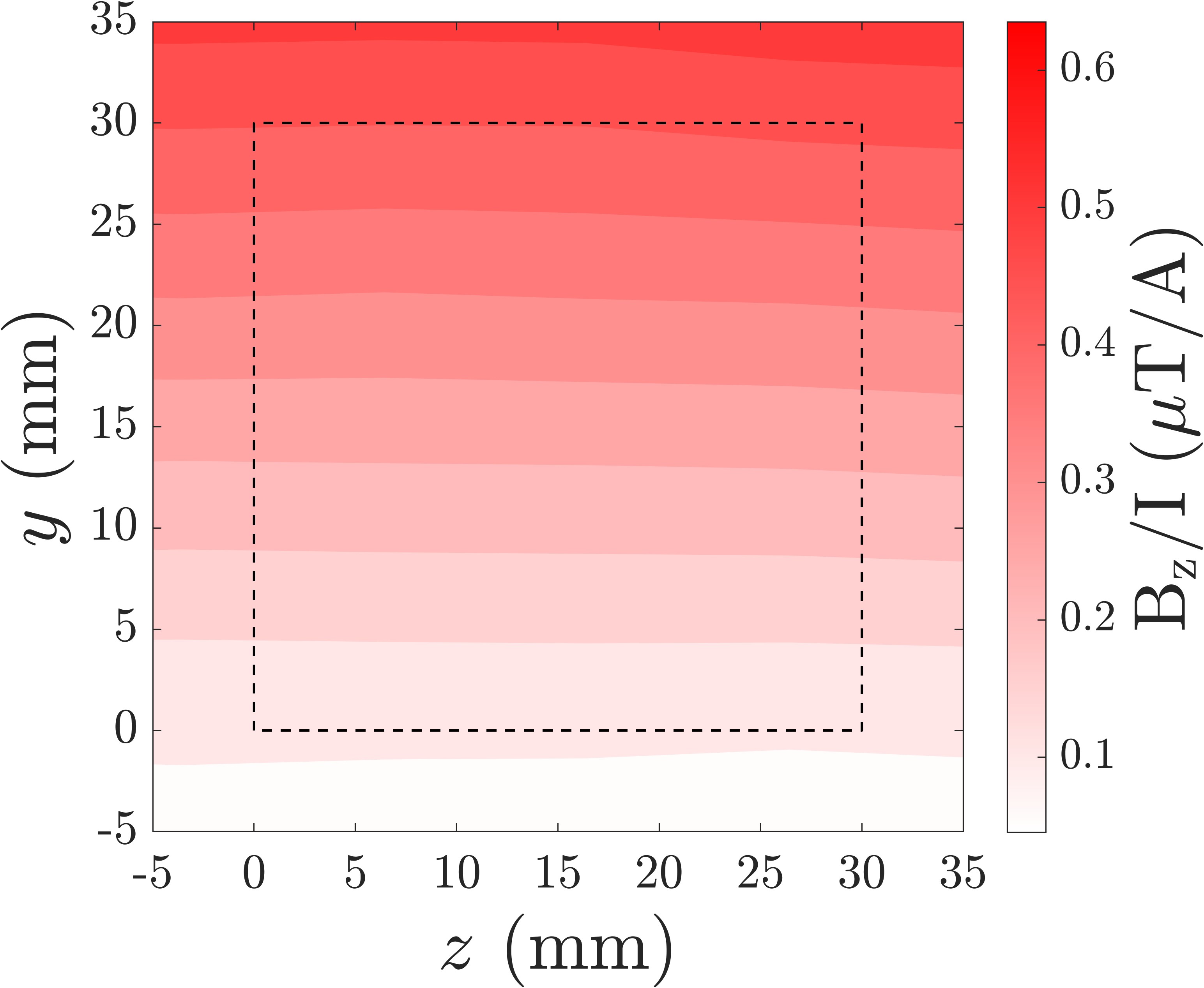} \\
         (a) & (b) \\
    \end{tabular}
\caption{Colourmap of $B_z/\mathrm{I}$ along the $zy$-plane generated by the $\mathbf{B}_{2,-1}$ coil, presented (a) at $x=0$~mm and (b) at $x=30$~mm. The dashed square encloses part of the target region between $0 - 30$~mm in along both the $y$ and $z$ coordinates.}
\label{fig.dBz colormap}
\end{center}
\end{figure}
The background noise limits mean that an equivalent maximum deviation calculation for the $\mathbf{B}_{2,1}$ coil is infeasible. This is because the noise contribution would be significantly larger when examining the difference between readings close to $y=0$, where the $B_z$ magnitude is small. Instead, in Figure~\ref{fig.dBz colormap}, we present the gradient profile across two slices of the $zy$-plane at $x=0$~mm in the centre of the target region and at the edge of the target region at $x=30$~mm. Performing fits to $B_z\left(y\right)=B_1 + B_2 y$ within the part of the target region bounded within dashed squares in Figure~\ref{fig.dBz colormap}, we find that the target field gradient reduces by $\left(2\pm1\right)$\% from the centre slice to the edge slice. Thus, the measured error in the $\mathbf{B}_{2,-1}$ coil is consistent with the error in the $\mathbf{B}_{1,0}$ coil presented in the analysis above. However, like previously, further measurements should be performed in a shielded environment to minimise measurement noise.

\section{Conclusions}
This paper presents the development and validation of a system designed to stabilise magnetic field variations experienced by an inertial sensor based on atom interferometry. The system utilises coils to create a uniform axial magnetic field, stronger than Earth's magnetic field, housed on bi-planar rectangular panels. Additionally, a second set of panels is used to generate a transverse gradient in the axial field along the atomic path. The system is retrofitted to the sensor to not only enhance its existing capability, but also validate the coil design method and investigate its practical integration.

We began by deriving a coil design model that relates the magnetic field inside a pair of rectangular planes to modes of surface current bound to the planar surfaces. We then used symmetry arguments to determine which of these modes relate to the required target magnetic fields in the quantum sensor. We found the optimal contributions of the modes to generate each target magnetic field profile using least-squares optimisation, regularised by the curvature of the current flows. We showed how the surface current can be approximated using sets of discrete wires, which are then converted into grooves in CNC-routed acrylic panels. The panels were wound by-hand using copper wire and mounted to the sensor's frame for experimental validation.

The experimentally measured magnetic field profiles were shown to have excellent spatial agreement to their desired forms along the $y$-axis of the target region of length $60$~mm, which contains the atomic path. The experimentally measured uniform axial field is measured to deviate by less than $1.30$\% throughout this region, with the limiting factor being leakage of background electromagnetic noise into the measurement. Future measurements could be repeated in a shielded environment, although care would need to be taken that the coils do not electromagnetically couple to any external passive shielding. While the quality of the magnetic fields measured experimentally is broadly equivalent to that theoretically predicted, the measured fields are slightly weaker due to positioning errors when placing the coils on the sensor. The field quality was preserved, despite this scaling, because the designs were intentionally highly regularised to improve their power efficiency, which also made the target fields less sensitive to misalignment when affixing the panels onto the rig. This is particularly useful in a large-scale system and if the device is operated in an environment where the coils may be displaced, e.g. due to hostiles. Errors would increase in scenarios where the target region needs to be larger relative to the coil size or where the device size reduces. This may necessitate more complex, less regularised coil designs with a more severe positional tolerance requirement. 

More broadly, CNC routing is shown to be an effective manufacturing route to emulate surface current patterns using discrete wires. In the future, we will explore using multi-layered rigid PCBs to make these patterns. Rigid PCBs can be made to be very thin, do not require labour-intensive and time-consuming hand-winding, and enable different coil patterns to be housed together on different layers of a single board surface without distortion. However, PCBs are also significantly more expensive to make and are harder to repair and maintain, and so may not be suitable for retrofitting field-deployable devices or those designed for hostile environments. In cases where CNC-boards are preferred, the design of the boards should be directly integrated into the mounting frame design, for example, by including specialised clips and mechanical connectors to ensure positional accuracy. As previously discussed, the mounting frame of the sensor, contains ample space along the $yz$-plane to retrofit further coil systems, e.g. to null uniform $B_x$ and $B_y$ transverse fields. If more target fields need to be generated using the same number of coil planes, then one could use a single pair of active cancellation coils to generate multiple magnetic field harmonics, i.e. by using two different relative current values between the planes~\citep{Holmes2019}, at the cost of reduced harmonic quality.

The next steps for implementing the coil system will involve in-situ testing by examining the sensor's performance with and without the bi-planar magnetic stabilisation coils. This could be implemented by considering the deviation from a known dead-reckoning with and without active magnetic field stabilisation to suppress magnetically-induced errors. In situ operation of the coils will require magnetometers to be placed around the edge of the target region, so that the target field can be decomposed into spherical harmonics, following the method in~\cite{10177829}. Computer-controlled coil drivers can then update the applied currents to generate equal and opposite fields to the measured background. If high frequency magnetic field stabilisation is required, the design methodology could be modified to include constraints to minimise the inductance of the coils~\citep{10.1088/0022-3735/21/10/008} to enable faster switching. Similar coil systems manufactured using CNC routing may also be retrofitted to existing atomic experiments moving from the laboratory environment to portable devices, including as active magnetic field control systems in a Cube-SAT package for space-based atom interferometry~\citep{Trimeche_2019} or coupled to passive magnetic shielding inside lightweight magnetically shielded rooms~\citep{10.1038/s41598-022-17346-1} for cardiac and brain imaging using atomic magnetometers.

%%%%%%%%%%%%%%%%%%%%%%%%%%%%%%%% END OF ARTICLE %%%%%%%%%%%%%%%%%%%%%%%%%%%%

\section*{Acknowledgements}
We acknowledge the support of the UK Quantum Technology Hub Sensors and Timing (EP/T001046/1) and from the Defence Science and Technology Laboratory (DSTL). \\

We also acknowledge Sionnach Devlin, a Senior Technician in the School of Physics \& Astronomy workshop, for his assistance in the design and manufacture of the panels.

\bibliographystyle{elsarticle-harv}
\bibliography{bibliography}
%%%%%%%%%%%%%%%%%%%%%%%%%%%%%% - APPENDIX - %%%%%%%%%%%%%%%%%%%%%%%%%%%%%%%%
\appendix
\section{Mathematical Appendix} \label{App.maths}
\subsection{Fourier transforms}
Substituting equation~\eqref{eq.streamyrect} into equation~\eqref{eq.jsrect}, we can determine the current density components in terms of the Fourier modes,
\begin{multline}\label{eq.FTJXpose}
    J_{x}\left(x',y'\right) = \left[\mathcal{H}\left(x'-\frac{L_{cx}}{2}\right)-\mathcal{H}\left(x'+\frac{L_{cx}}{2}\right)\right] \\
    \left[\mathcal{H}\left(y'-\frac{L_{cy}}{2}\right)-\mathcal{H}\left(y'+\frac{L_{cy}}{2}\right)\right] \times \\
    \sum_{n_x=1}^{N_x}\sum_{n_y=1}^{N_y}\ n_y\sqrt{\frac{L_{cx}}{L_{cy}}} A_{{n_x}{n_y}} \\
    \sin\left(\frac{n_x\pi\left(x'-L_{cx}/2\right)}{L_{cx}}\right)\cos\left(\frac{n_y\pi\left(y'-L_{cy}/2\right)}{L_{cy}}\right),
\end{multline}
\begin{multline}\label{eq.FTJYpose}
    J_{y}\left(x',y'\right) = -\left[\mathcal{H}\left(x'-\frac{L_{cx}}{2}\right)-\mathcal{H}\left(x'+\frac{L_{cx}}{2}\right)\right] \\
    \left[\mathcal{H}\left(y'-\frac{L_{cy}}{2}\right)-\mathcal{H}\left(y'+\frac{L_{cy}}{2}\right)\right] \times \\
    \sum_{n_x=1}^{N_x}\sum_{n_y=1}^{N_y}\ n_x\sqrt{\frac{L_{cy}}{L_{cx}}} A_{{n_x}{n_y}} \\
    \cos\left(\frac{n_x\pi\left(x'-L_{cx}/2\right)}{L_{cx}}\right)\sin\left(\frac{n_y\pi\left(y'-L_{cy}/2\right)}{L_{cy}}\right).
\end{multline}

To solve for the magnetic field in terms of these modes, we must calculate the Fourier transform, \eqref{eq.FT}, of the current density. We will find it useful to separate the Fourier transform into separate terms which depend exclusively on the $x'$ and $y'$ variables. For example, the Fourier transform of the $J_{x}\left(x',y'\right)$ component, \eqref{eq.FTJXpose}, is
\begin{equation} \label{eq.FTx}
        \tilde{J}_{x}\left(k_x,k_y\right) = \sum_{n_x=1}^{N_x}\sum_{n_y=1}^{N_y}\ \frac{n_y}{\pi^2} \sqrt{\frac{L_{cx}}{L_{cy}}} A_{{n_x}{n_y}} f_x\left(k_x\right) \overline{f}_x\left(k_y\right),
\end{equation}
where
\begin{align}\label{eq.fxkx}
     f_{x}\left(k_x\right) = \int_0^{\frac{L_{cx}}{2}} \mathrm{d}x' \ \begin{bmatrix}
    \cos\left(k_{x}x'\right) \\ -i\sin\left(k_{x}x'\right)
\end{bmatrix} \sin{\left(\frac{n_x \pi (x' - {L_{cx}}/{2})}{L_{cx}}\right)},
\end{align}
\begin{align}\label{eq.fxky}
     \overline{f}_{x}\left(k_y\right) = \int_0^{\frac{L_{cy}}{2}} \mathrm{d}y' \ \begin{bmatrix}
    -i\sin\left(k_{y}y'\right) \\ \cos\left(k_{y}y'\right)
\end{bmatrix} \cos{\left(\frac{n_y \pi (y' - {L_{cy}}/{2})}{L_{cy}}\right)},
\end{align}
and the brackets, $\left[\begin{smallmatrix}
  \mathbb{A}\\
  \mathbb{B}
\end{smallmatrix}\right]$, denote cases where the mode order, $n=\left(n_x,n_y\right)$, is odd or even $\mathbb{A} \in 2n - 1$, $\mathbb{B} \in 2n$ where $n \in \mathbb{Z^{+}}$.

These integrals may be solved by decomposing the sinusoidal variations into constituent frequency terms for odd and even orders separately. For example, the sinusoidal variation in equation~\eqref{eq.fxkx}, for odd modes. is
\begin{align}
\sin{\left(\frac{n_x \pi (x' - {L_{cx}}/{2})}{L_{cx}}\right)} &= -\sin\left( \frac{n_x \pi}{2} \right) \cos\left( \frac{n_x \pi}{L_{cx}}x' \right) \nonumber \\ &= -\cos\left( \frac{n_x \pi}{L_{cx}}x' \right).
\end{align}
Thus, equation~\eqref{eq.fxkx} for odd modes is simply~\citep{mathsbook},
\begin{align}\label{eq.fx-kx-odd}
     f_{x}\left(k_x\right)^{\mathrm{odd}} &= -\int_0^{\frac{L_{cx}}{2}} \mathrm{d}x' \ \cos\left(k_{x}x'\right) \cos\left( \frac{n_x \pi}{L_{cx}}x' \right) \nonumber \\
     &= \frac{n_x\pi L_{cx}}{k_{x}^{2}L_{cx}^{2} - n_x^{2}\pi^{2}} \cos\left(\frac{k_{x}L_{cx}}{2}\right).
\end{align}
Repeating this approach, we find that the Fourier transforms in equations~\eqref{eq.fxkx}~and~\eqref{eq.fxky} are
\begin{equation}
    f_x(k_x) = \frac{n_x \pi L_{cx}}{k_{x}^{2}L_{cx}^{2} - {n_x}^{2}\pi^{2}} 
    \begin{bmatrix}
    \cos\left({k_{x}L_{cx}}/{2}\right)  \\
    -i\sin\left({k_{x}L_{cx}}/{2}\right)  
     \end{bmatrix},
\end{equation}
\begin{equation}
    \overline{f}_x(k_y) = \frac{k_y L_{cy}^2}{k_{y}^{2}L_{cy}^{2} - {n_y}^{2}\pi^{2}} 
    \begin{bmatrix}
    i\cos\left({k_{y}L_{cy}}/{2}\right)  \\
    \sin\left({k_{y}L_{cy}}/{2}\right)  
     \end{bmatrix}.
\end{equation}
Finally, by association, we note that the Fourier transform of the $J_{y}\left(x',y'\right)$ component, \eqref{eq.FTJYpose}, is similar but with opposite correspondences, such that
\begin{equation} \label{eq.FTy}
        \tilde{J}_{y}\left(k_x,k_y\right) = -\sum_{n_x=1}^{N_x}\sum_{n_y=1}^{N_y}\ \frac{n_x}{\pi^2} \sqrt{\frac{L_{cy}}{L_{cx}}} A_{{n_x}{n_y}} \overline{f}_y\left(k_x\right) f_y\left(k_y\right),
\end{equation}
where
\begin{align}
    f_y(k_y) = \frac{n_y \pi L_{cy}}{k_{y}^{2}L_{cy}^{2} - {n_y}^{2}\pi^{2}} 
    \begin{bmatrix}
    \cos\left({k_{y}L_{cy}}/{2}\right)  \\
    -i\sin\left({k_{y}L_{cy}}/{2}\right)  
     \end{bmatrix},
\end{align}
\begin{align}
    \overline{f}_y(k_x) = \frac{k_x L_{cx}^2}{k_{x}^{2}L_{cx}^{2} - {n_x}^{2}\pi^{2}} 
    \begin{bmatrix}
    i\cos\left({k_{x}L_{cx}}/{2}\right)  \\
    \sin\left({k_{x}L_{cx}}/{2}\right)  
     \end{bmatrix}.
\end{align}

\subsection{Master equations}
Substituting equations~\eqref{eq.FTx}~and~\eqref{eq.FTy} into equations~\eqref{eq.Bx.rel.2}--\eqref{eq.Bz.rel.1} and summing over the contributions of two bi-planes at positions $z={\pm}z_c$, we find
\begin{strip}
\begin{multline} \label{Bx_pseudo_final}
    B_{x}\left(x,y,z\right) = \frac{4  L_{cx}^{3/2} L_{cy}^{3/2} \mu_{0}}{\pi^2}   \sum_{n_x=1}^{N_x}\sum_{n_y=1}^{N_y}\ A_{{n_x}{n_y}} \int_{0}^{\infty}\mathrm{d}k_x \ \int_{0}^{\infty}\mathrm{d}k_y \  \frac{n_x n_y k_x e^{-\sqrt{k_{x}^{2} + k_{y}^{2}}~z_{c}}}{(k_x^{2}L_{cx}^{2} - {n_x}^{2}\pi^{2})(k_y^{2}L_{cy}^{2} - {n_y}^{2}\pi^{2})} \ \times \\ 
    \begin{bmatrix}
     \cos\left({k_x L_{cx}}/2\right) \sin\left(k_x x\right)   \\
     -\sin\left({k_x L_{cx}}/2\right) \cos\left(k_x x\right)
    \end{bmatrix}
    \begin{bmatrix}
     \cos\left({k_y L_{cy}}/2\right) \cos\left(k_y y\right)   \\
     \sin\left({k_y L_{cy}}/2\right) \sin\left(k_y y\right)
    \end{bmatrix} 
    \begin{Bmatrix}
     \sinh\left({\sqrt{k_{x}^{2} + k_{y}^{2}}}~z\right)  \\
     \cosh\left({\sqrt{k_{x}^{2} + k_{y}^{2}}}~z\right)
    \end{Bmatrix},    
\end{multline}

\begin{multline} \label{By_pseudo_final}
    B_{y}\left(x,y,z\right) = \frac{4  L_{cx}^{3/2} L_{cy}^{3/2} \mu_{0}}{\pi^2} \sum_{n_x=1}^{N_x}\sum_{n_y=1}^{N_y}\ A_{{n_x}{n_y}} \int_{0}^{\infty}\mathrm{d}k_x \ \int_{0}^{\infty}\mathrm{d}k_y \ \frac{n_x n_y k_y e^{-\sqrt{k_{x}^{2} + k_{y}^{2}}~z_{c}}}{(k_x^{2}L_{cx}^{2} - {n_x}^{2}\pi^{2}) (k_y^{2}L_{cx}^{2} - {n_y}^{2}\pi^{2})} \ \times   \\  
    \begin{bmatrix}
     \cos\left({k_x L_{cx}}/2\right) \cos\left(k_x x\right)   \\
     \sin\left({k_x L_{cx}}/2\right) \sin\left(k_x x\right)
    \end{bmatrix}  
    \begin{bmatrix}
     \cos\left({k_y L_{cy}}/2\right) \sin\left(k_y y\right)   \\
     -\sin\left({k_y L_{cy}}/2\right) \cos\left(k_y y\right)
    \end{bmatrix} 
     \begin{Bmatrix}
     \sinh\left({\sqrt{k_{x}^{2} + k_{y}^{2}}}~z\right)  \\
     \cosh\left({\sqrt{k_{x}^{2} + k_{y}^{2}}}~z\right)
    \end{Bmatrix}
     ,
\end{multline}

\begin{multline} \label{Bz_pseudo_final}
    B_{z}\left(x,y,z\right) = -\frac{4  L_{cx}^{3/2} L_{cy}^{3/2} \mu_{0}}{\pi^2} \sum_{n_x=1}^{N_x}\sum_{n_y=1}^{N_y}\ A_{{n_x}{n_y}} \int_{0}^{\infty}\mathrm{d}k_x \ \int_{0}^{\infty}\mathrm{d}k_y \ \frac{n_x n_y \sqrt{k_{x}^{2} + k_{y}^{2}} e^{-\sqrt{k_{x}^{2} + k_{y}^{2}}~z_{c}}}{(k_x^{2}L_{cx}^{2} - {n_x}^{2}\pi^{2}) (k_y^{2}L_{cy}^{2} - {n_y}^{2}\pi^{2})} \times \\
    \begin{bmatrix}
     \cos\left({k_x L_{cx}}/2\right) \cos\left(k_x x\right)   \\
     \sin\left({k_x L_{cx}}/2\right) \sin\left(k_x x\right)
    \end{bmatrix}    
    \begin{bmatrix}
     \cos\left({k_y L_{cy}}/2\right) \cos\left(k_y y\right)   \\
     \sin\left({k_y L_{cy}}/2\right) \sin\left(k_y y\right)
    \end{bmatrix} 
     \begin{Bmatrix}
     \cosh{\left(\sqrt{k_{x}^{2} + k_{y}^{2}}~z\right)}  \\
     \sinh{\left(\sqrt{k_{x}^{2} + k_{y}^{2}}~z\right)}
    \end{Bmatrix},
\end{multline}
\end{strip}
where $\left\{\begin{smallmatrix}
  +\\
  - \end{smallmatrix}
\right\}$, denotes the cases where the coil currents on the pair of planes are equal, $(+)$, or opposite, $(-)$, and, like previously, the brackets, $\left[\begin{smallmatrix}
  \mathbb{A}\\
  \mathbb{B}
\end{smallmatrix}\right]$, denote cases where the mode order, $n=\left(n_x,n_y\right)$, is odd or even $\mathbb{A} \in 2n - 1$, $\mathbb{B} \in 2n$ where $n \in \mathbb{Z^{+}}$.

\subsection{Curvature}
Let us derive the curvature of current flows on the surface of a rectangular plane in terms of the rectangular planar streamfunction, \eqref{eq.streamyrect}. The scalar Laplacian of the streamfunction is
\begin{align}\label{eq.laprect}
    \mathbf{\nabla}^2\varphi\left(x',y'\right) &= -\pi^2\left(\left(\frac{n_x}{L_{cx}}\right)^2 + \left(\frac{{n_y}^2}{L_{cy}}\right)\right)\varphi\left(x',y'\right),
\end{align}
We can then insert equation~\eqref{eq.laprect} into equation~\eqref{eq.crv} and integrate using double angle formulae to find
\begin{align}
    C &= \frac{\gamma\pi^2}{4} \sum_{n_x=1}^{N_x}\sum_{n_y=1}^{N_y}\ A_{n_x{n_y}}^2\left({n_x}^2\frac{L_{cy}}{L_{cx}} + {n_y}^2\frac{L_{cx}}{L_{cy}}\right)^2.
\end{align}

\end{document}